\magnification \magstep1
\raggedbottom
\openup 4\jot
\voffset6truemm
\def\cstok#1{\leavevmode\thinspace\hbox{\vrule\vtop{\vbox{\hrule\kern1pt
\hbox{\vphantom{\tt/}\thinspace{\tt#1}\thinspace}}
\kern1pt\hrule}\vrule}\thinspace}
\leftline {\bf ON THE OCCURRENCE OF MASS IN FIELD THEORY}
\vskip 0.3cm
\leftline {\bf Giampiero Esposito}
\vskip 0.3cm
\noindent
{\it INFN, Sezione di Napoli,
Complesso Universitario di Monte S. Angelo, Via Cintia, Edificio N',
80126 Napoli, Italy; e-mail: giampiero.esposito@na.infn.it}
\vskip 0.3cm
\noindent
{\it Dipartimento di Scienze Fisiche,
Universit\`a di Napoli ``Federico II'',
Complesso Universitario di Monte S. Angelo, Via Cintia, Edificio N',
80126 Napoli, Italy}
\vskip 1cm
\leftline {Received 19 April 2002, revised 10 June 2002}
\vskip 1cm
\noindent
This paper proves that it is possible to build a Lagrangian
for quantum electrodynamics which makes it explicit that the photon
mass is eventually set to
zero in the physical part on observational ground.
Gauge independence is achieved upon considering the joint effect of
gauge-averaging term and ghost fields.
It remains possible to obtain a counterterm Lagrangian where the only
non-gauge-invariant term is proportional to the squared divergence
of the potential,
while the photon propagator in momentum space falls
off like $k^{-2}$ at large $k$ which indeed agrees
with perturbative renormalizability. The resulting
radiative corrections to the Coulomb potential in QED are also
shown to be gauge-independent. The experience acquired
with quantum electrodynamics is used to investigate properties
and problems of the extension of such ideas to non-Abelian
gauge theories.
\vskip 2cm
\leftline {\bf 1. INTRODUCTION}
\vskip 1cm
\noindent
A key task of theoretical physics has been always the description of a
wide variety of natural phenomena within a unified conceptual framework,
where they can all be derived from a few basic principles which have
been carefully tested against observation. The development of local
or non-local field theories, the investigation of perturbative and
nonperturbative properties, and the construction of gauge theories of
fundamental interactions provide good examples of how such a task can
be accomplished. Moreover, when a commonly accepted model remains
unproven for a long time, the theoretical physicist has to perform a
careful assessment of the ideas leading to such
a prediction, and he/she is
expected to find either an independent way to confirm it, or an alternative
way to understand the phenomenon.

Within this framework, it is the aim of our paper to reconsider a
longstanding problem in particle physics and field theory, i.e. the
generation of mass in gauge theories of fundamental interactions. Although
the Higgs mechanism provides a well understood theoretical model for the
generation of mass,$^{(1)}$ the analysis
of alternative models appears necessary
for at least a fundamental reason: no conclusive evidence on the existence
of the Higgs field is available as yet. At present one can only say that,
from the precision measurements of the mass of the $W$ boson and the
effective leptonic weak mixing angle at the $Z$-boson resonance, one
finds a $95$ per cent confidence level upper bound on the Higgs-boson mass
given by $M_{H} < 188$ GeV.$^{(2)}$ For
example, in the Weinberg--Salam model,$^{(3-5)}$ the Lagrangian density
$\cal L$ (hereafter we omit the word ``density'' for simplicity)
contains five terms describing gauge bosons, the coupling of gauge
bosons to scalars, the coupling of gauge bosons to left-handed
and right-handed fermions, and the
gauge-invariant interaction among scalars and fermions, respectively.
In particular, the coupling of gauge bosons to scalars is described
by the term
$$
{\cal L}_{GB-S}=(D^{\mu}\phi)^{\dagger}D_{\mu}\phi
-V(\phi^{\dagger}\phi),
\eqno (1.1)
$$
where $\phi$ is a Higgs field and the gauge-covariant derivative reads
$$
D_{\mu} \equiv \partial_{\mu}+ig\sum_{k=1}^{3}W_{\mu}^{k}\tau_{k}
+ig' W_{\mu}^{0}\tau_{0}.
\eqno (1.2)
$$
With a standard notation, $W_{\mu}^{k}$ are the $SU(2)$ gauge fields
with associated generators $\tau_{k}$, while $W_{\mu}^{0}$ is the
$U(1)$ gauge field with generator $\tau_{0}={1\over 2}
\pmatrix{1&0 \cr 0&1 \cr}$. In the unitary gauge, the Higgs field is
expressed by the ``column vector''
$\phi=\pmatrix{0 \cr {\widetilde \rho} \cr}$,
and after writing the transformation ($\theta_{w}$ being the Weinberg
angle)
$$
\pmatrix{W_{\mu}^{3} \cr W_{\mu}^{0} \cr}=
\pmatrix{\cos \theta_{w} & \sin \theta_{w} \cr -\sin \theta_{w} &
\cos \theta_{w} \cr}
\pmatrix{Z_{\mu} \cr A_{\mu} \cr},
\eqno (1.3)
$$
the kinetic term in Eq. (1.1) reads eventually
$$
(D^{\mu}\phi)^{\dagger}D_{\mu}\phi = {g^{2}\over 4}\Bigr(W_{\mu}^{1}
W_{1}^{\mu}+W_{\mu}^{2}W_{2}^{\mu}\Bigr){\widetilde \rho}^{2}
+ {g^{2}\over 4}{Z^{\mu}Z_{\mu}{\widetilde \rho}^{2}
\over \cos^{2}\theta_{w}}.
\eqno (1.4)
$$
Thus, the vector mesons $W_{+} , W_{-}$ and $Z$
are found to have square masses ${1\over 2}g^{2}{\widetilde \rho}^{2},
{1\over 2}g^{2}{\widetilde \rho}^{2}$ and
${1\over 2 \cos^{2} \theta_{w}}g^{2}{\widetilde \rho}^{2}$,
respectively. From the known experimental value of the Weinberg angle,
one then finds at tree level masses
$m_{W}$ and $m_{Z}$ of order $80$ GeV and $90$ GeV,
respectively. Nevertheless, since the Higgs field remains unobserved,
we are led to ask ourselves whether the fundamental principles of
quantum field theory make it possible to fit the experimental data
without having to assume the existence of a Higgs field.

Motivated by this outstanding problem,
Secs. 2 and 3 study a new class of gauge-averaging
functionals in the path integral for bosonic gauge theories,
and other original results
are derived in Secs. 4--8, which are devoted to photon
propagators in quantum electrodynamics; perturbative renormalization of
a QED model where the mass of the photon is set to zero only on observational
ground at a later stage; radiative corrections in QED;
mass terms for vector mesons in non-Abelian gauge theory.
Concluding remarks and open problems are presented in Sec. 9.
\vskip 2cm
\leftline {\bf 2. GAUGE-AVERAGING FUNCTIONALS AND GAUGE-FIELD}
\leftline {\bf OPERATORS}
\vskip 1cm
At this stage, the fundamental point in our investigation is the need to
recall a well known property of all gauge theories: since an invariance
group is present, the operator obtained from second functional
derivatives $S_{,ij}$ of the classical action $S$ is not invertible.
To obtain an invertible operator on field disturbances one has to add
to $S_{,ij}$ a term obtained from the generators of infinitesimal
gauge transformations and their adjoints.$^{(6)}$ In the corresponding
quantum theory, the counterpart of this construction is the addition of
a gauge-averaging (also called, more frequently, gauge-breaking or
gauge-fixing) term to the original Lagrangian $\cal L$.$^{(7)}$ The
resulting Lagrangian leads to well defined functional determinants in the
one-loop semiclassical theory and is part of the path-integral prescription
for gauge theories, aimed at avoiding a ``summation'' over gauge-equivalent
field configurations for the out-in amplitude. In other words, the key
idea which inspires our model is as follows:
a gauge-invariant Lagrangian is very elegant, but what one really
needs is instead a Lagrangian leading to an invertible operator on
gauge fields,$^{(6-8)}$ with the associated gauge-breaking term and
ghost fields.$^{(6-12)}$ This is invariant under BRST transformations,
which express the most important symmetry of modern quantum
field theory. [We will find that the resulting theory cannot truly
``generate'' mass, but acquires technical tools for describing
its occurrence]

Bearing in mind these properties, let us consider for simplicity
the Lagrangian for Euclidean Maxwell theory via path integrals
(here we write only the part involving the potential):
$$
{\cal L}={1\over 4}F_{\mu \nu}F^{\mu \nu}
+{1\over 2\alpha}[\Phi(A)]^{2}.
\eqno (2.1)
$$
With a standard notation, $F_{\mu \nu}$ is the electromagnetic field strength
that contributes the non-invertible operator ($R_{\mu \nu}$ being the Ricci
tensor of the background with metric $g$)
$$
-g_{\mu \nu} \cstok{\ } +\nabla_{\mu}\nabla_{\nu}+R_{\mu \nu}
$$
acting on the potential (with $\cstok{\ } \equiv \nabla^{\mu}
\nabla_{\mu}=g^{\mu \nu}\nabla_{\mu}\nabla_{\nu}$,
and $\nabla$ the Levi-Civita
connection on space-time). Moreover, $\alpha$ is a dimensionless
parameter, and $\Phi$ is the gauge-averaging functional
$$
\Phi: A \in \left \{ A_{\mu}dx^{\mu} \right \} \rightarrow
\Phi(A) \in {\bf R}.
$$
The potential $A$ is mapped into the real number $\Phi(A)$ via the action
of $\Phi$ in a way here expressed in the form
$$
\Phi(A) \equiv T^{\mu}A_{\mu}=g^{\mu \nu}T_{\nu}A_{\mu}.
\eqno (2.2)
$$
In a local formulation, $T^{\mu}=\nabla^{\mu}$ leads to the Lorenz
gauge,$^{(13)}$ with the associated gauge-field operator
$$
P_{\mu \nu}=-g_{\mu \nu}\cstok{\ } + \left(1-{1\over \alpha}\right)
\nabla_{\mu}\nabla_{\nu}+R_{\mu \nu},
\eqno (2.3)
$$
which becomes of Laplace type (in a Euclidean framework) when $\alpha$
is set equal to $1$ (this is the Feynman choice for $\alpha$).
\vskip 2cm
\leftline {\bf 3. ROLE OF $\gamma$-MATRICES}
\vskip 1cm
Independently of being able to find an alternative to the Higgs
mechanism in non-Abelian theories, we would now like it to understand
whether mass terms can be considered with the help of a suitable
formulation of the process of gauge-averaging
in the path integral, while making sure that such masses are
unaffected by any particular choice of gauge parameters.
In the simpler case of Maxwell theory, we should
find how a term proportional to $A_{\mu}A^{\mu}$ can be obtained.
Indeed, in four dimensions, one can exploit the identity
$$
A_{\mu}A^{\mu}=g^{\mu \nu}A_{\mu} A_{\nu}
={1\over 4}{\rm Tr}(\gamma^{\mu}\gamma^{\nu})A_{\mu}A_{\nu},
$$
where $g^{\mu \nu}$ are the contravariant components of the metric
tensor.

We have therefore looked, in a first moment, for a gauge-fixing condition
combining the effect of Lorenz
gauge and $\gamma$-matrices.
However, one cannot simply add the derivatives of $A_{\mu}$ in the Lorenz
gauge and $\gamma^{\mu}$ terms, since the latter are four-vectors with
components given by $4 \times 4$ matrices. The only well defined
operation on such objects is the one giving rise
to the matrix (here $i,j$
are matrix indices ranging from $1$ through $4$)
$$
\Phi_{i}^{\; j}(A) \equiv
\Bigr(\delta_{i}^{\; j}\partial^{\mu}
+\beta (\gamma^{\mu})_{i}^{\; j}\Bigr)A_{\mu}(x),
\eqno (3.1)
$$
where we use a notation that makes it explicit how to add correctly
partial derivatives and $\gamma$-matrix contributions, and
the parameter $\beta$ is now introduced to ensure that all terms in
$\Phi_{i}^{\; j}$ have the same dimension
(i.e. $\beta$ has dimension [length]$^{-1}$).
There is only one coefficient, $\beta$, since only
one potential $A_{\mu}$ is available for contraction with
$\gamma^{\mu}$ in the Abelian case.
The resulting gauge-averaging term in the path integral for quantum
electrodynamics is taken to be (with $\alpha$ a real parameter)
$$
{\Phi^{2}(A)\over 2\alpha}
={1\over 2\alpha}\Phi_{i}^{\; j}(A) \Omega_{j}^{\; k}
\Phi_{k}^{\; i}(A),
$$
where, having defined the symmetric matrix
$$
\Omega_{j}^{\; k} \equiv {1\over 4}
\delta_{j}^{\; k} ,
\eqno (3.2)
$$
one finds
$$ \eqalignno{
\; & \Phi_{i}^{\; j}(A) \Omega_{j}^{\; k}
\Phi_{k}^{\; i}(A)
={1\over 4} \Bigr[4(\partial^{\mu}A_{\mu})(\partial^{\nu}A_{\nu})
+\beta (\partial^{\mu}A_{\mu})(\gamma^{\nu})_{i}^{\; i}A_{\nu}
+\beta (\gamma^{\mu})_{i}^{\; i}A_{\mu}\partial^{\nu}A_{\nu} \cr
&+ \beta^{2}(\gamma^{\mu})_{i}^{\; j}
(\gamma^{\nu})_{j}^{\; i}
A_{\mu}A_{\nu}\Bigr]
=(\partial^{\mu}A_{\mu})(\partial^{\nu}A_{\nu})
+\beta^{2} A_{\mu}A^{\mu},
&(3.3)\cr}
$$
since the $\gamma$-matrices are traceless,
and ${\rm Tr}(\gamma^{\mu}\gamma^{\nu})
=4 g^{\mu \nu}$. When $\beta$ is set to zero, this reduces to the familiar
Lorenz gauge-averaging term, which is why the numerical factors have been
chosen as in the definition (3.2).

It should be stressed that the matrix (3.1) is a
tool to express in a concise and elegant form the gauge-averaging
term ${\Phi^{2}(A)\over 2\alpha}$ in the full action, but {\it our
gauge-averaging functional for} QED {\it is not a matrix} and is
equal to
$$
\Phi(A) \equiv \sqrt{\Phi_{i}^{\; j}(A)\Omega_{j}^{\; k}
\Phi_{k}^{\; i}(A)}.
\eqno (3.4)
$$
One cannot regard $\Phi_{i}^{\; j}(A)$ itself as a
gauge-averaging functional, since otherwise one would get $16$
supplementary conditions which are totally extraneous to the
quantum (as well as classical) theory and make it over-constrained.
To make sure that the equation $\Phi(A)=\zeta$ admits a solution
for all real $\zeta$, we have to require that the right-hand
side of (3.3) should remain $\geq 0$ for all real $\beta$.
In Minkowski space-time, this is achieved with $A_{\mu}A^{\mu}>0$
and metric given by diag(1,-1,-1,-1), or with
$A_{\mu}A^{\mu}<0$ and metric given by diag(-1,1,1,1). In the
Euclidean regime, where the $\gamma$-matrices are
anti-Hermitian so that $g^{\mu \nu}=-{1\over 4}{\rm Tr}(\gamma^{\mu}
\gamma^{\nu})$, one has then to take $\beta$ pure imaginary.

Note also that, on denoting by $I$ the $4 \times 4$ identity matrix,
the operator $D^{\mu} \equiv I \partial^{\mu}+\beta \gamma^{\mu}$ in
Eq. (3.1) bears apparently some resemblance
with the super-covariant derivative first
considered by Townsend$^{(14)}$ within the framework of supergravity in
anti-de Sitter space. However, the commutator $[D^{\mu},D^{\nu}]$ does not
vanish in Minkowski space-time, and is there equal to
$$
[D^{\mu},D^{\nu}]
=\beta^{2}(\gamma^{\mu}\gamma^{\nu}-\gamma^{\nu}\gamma^{\mu}).
\eqno (3.5)
$$
Thus, it is impossible to relate the commutator of these derivatives
to the space-time curvature, since the Riemann tensor of Minkowski
space-time vanishes. The gauge curvature does not help either, because,
on defining $\nabla_{\mu} \equiv \partial_{\mu}+q A_{\mu}$, one finds
$$
[\nabla_{\mu},\nabla_{\nu}]
=q (\partial_{\mu}A_{\nu}-\partial_{\nu}A_{\mu})
=q F_{\mu \nu},
\eqno (3.6)
$$
which does not yield (3.5) upon mapping $A$ into $\gamma$. This means
that no formal analogy can be actually
proposed between the operator occurring
in (3.1) and the Townsend construction$^{(14)}$ for supergravity in
anti-de Sitter.
\vskip 2cm
\leftline {\bf 4. PHOTON PROPAGATOR}
\vskip 1cm
As a first step towards quantization of non-Abelian theories$^{(15-17)}$
we now consider a simpler
but instructive problem, i.e. the photon propagator in the Euclidean
version of quantum electrodynamics with gauge-averaging functional (3.4).
In modern language, the path integral tells us that the photon propagator
is obtained by first evaluating the gauge-field operator $P_{\mu \nu}$
resulting from the particular choice of
gauge-averaging functional, then taking its
symbol $\sigma(P_{\mu \nu})$ and inverting such a symbol to find
$\sigma^{-1}(P_{\mu \nu})$ for which $\sigma \sigma^{-1}=\sigma^{-1}
\sigma=I$. The photon propagator reads eventually (cf. Ref. 18)
$$
\bigtriangleup^{\mu \nu}(x,y)=(2\pi)^{-4}\int_{\zeta} d^{4}k \;
\sigma^{-1}(P_{\mu \nu})e^{ik \cdot (x-y)}
\eqno (4.1)
$$
for some contour $\zeta$,
where $\sigma^{-1}(P_{\mu \nu}) \equiv {\widetilde \sigma}^{\mu \nu}$
should be thought of as carrying
contravariant indices, in agreement with the left-hand side.
In the light of (3.1) and (3.3), our gauge-field Lagrangian
(2.1) turns out to be, by virtue of gauge averaging,
$$
{\cal L}=\partial^{\mu}\rho_{\mu}+{1\over 2}A^{\mu}P_{\mu \nu}A^{\nu},
\eqno (4.2)
$$
where
$$
\rho_{\mu} \equiv -{1\over 2}A_{\nu}\partial^{\nu}A_{\mu}
+{1\over 2}A_{\nu}\partial_{\mu}A^{\nu}
+{1\over 2\alpha}A_{\mu}\partial^{\nu}A_{\nu},
\eqno (4.3)
$$
and
$$
P_{\mu \nu} \equiv g_{\mu \nu} \left[-\cstok{\ } +{\beta^{2}\over \alpha}
\right]+\left(1-{1\over \alpha}\right)\partial_{\mu}\partial_{\nu}.
\eqno (4.4)
$$
Of course, the term
$\rho_{\mu}$ only contributes to a total divergence and hence
does not affect the photon propagator, while the parameter $\alpha$ can be
set equal to $1$ (Feynman choice) so that calculations are simplified.
Thus, we can eventually obtain the gauge-field operator
$$
P_{\mu \nu}(\alpha=1)=g_{\mu \nu}\left(-\cstok{\ } +\beta^{2}\right).
\eqno (4.5)
$$
The symbol of (4.5), which results from Fourier analysis of our
translation-invariant operator, reads
$$
\sigma(P_{\mu \nu}(\alpha=1))=(k^{2}+\beta^{2})g_{\mu \nu},
\eqno (4.6)
$$
and hence our Euclidean photon propagator reads
$$
\bigtriangleup_{E}^{\mu \nu}(x,y)=(2\pi)^{-4}\int_{\Gamma}d^{4}k
{g^{\mu \nu}\over (k^{2}+\beta^{2})}e^{ik \cdot (x-y)},
\eqno (4.7)
$$
where the points $x$ and $y$ refer to the indices
$\mu$ and $\nu$, respectively.
Note that integration along the real axis for $k_{0},k_{1},k_{2},k_{3}$
avoids poles of the integrand, which are located at the complex points
for which $k^{2}=-\beta^{2}$. The choice $\alpha=1$ has led to Eq. (4.7)
which has the advantage of being very simple, but the correct
interpretation of the term $\beta^{2}$ will only be clear after reading
the following section.

In general, the gho\-st o\-pe\-ra\-tor is ob\-ta\-ined
by con\-trac\-ti\-on of func\-tio\-nal
de\-ri\-va\-ti\-ves of the ga\-uge-a\-ve\-ra\-ging
func\-tio\-nal wi\-th the in\-fi\-ni\-te\-si\-mal
ge\-ne\-ra\-to\-rs of ga\-uge tran\-sfor\-ma\-ti\-ons.$^{(6,7,17)}$
This leads to the $-\cstok{\ }$ operator in the Lorenz gauge, but we
are considering the non-linear gauge-averaging functional (3.4)
which therefore yields a more involved ghost operator $\cal P$,
whose action is given by
$$
{\cal P}: \varepsilon \rightarrow
\left[-{(\partial^{\nu}A_{\nu})\over \Phi(A)}\partial^{\mu}
\partial_{\mu}-{\beta^{2}A^{\mu}\over \Phi(A)}\partial_{\mu}
\right]\varepsilon .
\eqno (4.8)
$$
It reduces to the ghost operator in the Lorenz gauge after
imposing that photons are massless on observational ground
(see Sec. 5).
\vskip 2cm
\leftline {\bf 5. OUTLINE OF PERTURBATIVE RENORMALIZATION}
\vskip 1cm
A crucial question is now in order, i.e. how the perturbative renormalization
programme can be initiated with our choice (3.4) for the gauge-averaging
functional. For this purpose, we study the full Lagrangian
density $\cal L$ for spinor electrodynamics, including the
gauge-averaging term. Since we are aiming to split $\cal L$ into a
sum of physical and counterterm parts, we consider bare fields
denoted by the $B$ subscript and physical fields written without
such a subscript. To begin, for the gauge potential and the
spinor field we assume that renormalization can be performed
by considering the following relations:$^{(18)}$
$$
(A_{\mu})_{B}=\sqrt{z_{A}}\; A_{\mu}, \; \Longrightarrow
F_{B}^{\mu \nu}=\partial^{\mu}A_{B}^{\nu}
-\partial^{\nu}A_{B}^{\mu},
\eqno (5.1)
$$
$$
\psi_{B}=\sqrt{z_{\psi}} \; \psi,
\eqno (5.2)
$$
and similarly for mass, charge and gauge parameter respectively, i.e.
$$
m_{B}={z_{m}\over z_{\psi}}m,
\eqno (5.3)
$$
$$
e_{B}={z_{e}\over z_{\psi}\sqrt{z_{A}}}e,
\eqno (5.4)
$$
$$
\alpha_{B}={z_{A}\over z_{\alpha}}\alpha.
\eqno (5.5)
$$
Moreover, we also introduce, at the beginning of renormalization,
the equation
$$
\beta_{B}=\rho \beta,
\eqno (5.6)
$$
where $\rho$ can be fixed in due course (see below). It should be
stressed that multiplicative renormalizability of QED (i.e. the
renormalization relying upon the $z$-factors and $\rho$-factor)
with our gauge (3.4) is a conjecture at this stage, because
multiplicative renormalizability is not a universal property of
gauge theories independently of the gauge condition, but it will
be justified `a posteriori', once that the bare photon propagator
is obtained in (5.19), and bearing in mind what we said about
the ghost propagator following (4.8).

Now we are ready to
write the Lagrangian density in Minkowski space-time
(the part ${\cal L}_{\rm gh}$ involving Faddeev--Popov ghost fields
is not written explicitly, since it does not affect the
following calculations):
$$ \eqalignno{
{\cal L}-{\cal L}_{\rm gh}&= -{1\over 4}F_{B \; \mu \nu}F_{B}^{\mu \nu}
+{\overline \psi}_{B}\Bigr(i\gamma^{\mu}\partial_{\mu}
-e_{B}\gamma^{\mu}A_{B \; \mu}\Bigr)\psi_{B} \cr
&- m_{B}{\overline \psi}_{B}\psi_{B}-{1\over 2\alpha_{B}}
[\Phi(A_{B})]^{2},
&(5.7)\cr}
$$
where this general formula is here considered for $\Phi(A_{B})$
given by (3.4), with $\beta$ and
the potential $A_{\mu}$ replaced by
their renormalized values therein.
By virtue of (5.3)--(5.7) our Lagrangian density admits the split
$$
{\cal L}-{\cal L}_{\rm gh}
={\cal L}_{\rm {ph}}+{\cal L}_{\rm {ct}},
\eqno (5.8)
$$
where the physical part (also called basic) reads
$$ \eqalignno{
{\cal L}_{\rm ph}&= -{1\over 4}F_{\mu \nu}F^{\mu \nu}
+{\overline \psi}i\gamma^{\mu}\partial_{\mu}\psi
-e{\overline \psi}\gamma^{\mu}A_{\mu}\psi
-m{\overline \psi}\psi \cr
&- {1\over 2\alpha}(\partial^{\mu}A_{\mu})^{2}
-{\beta^{2}\over 2\alpha}A_{\mu}A^{\mu},
&(5.9)\cr}
$$
while the part involving counterterms is given by (cf. Ref. 18)
$$ \eqalignno{
{\cal L}_{\rm ct}&= -{1\over 4}(z_{A}-1)F_{\mu \nu}F^{\mu \nu}
+(z_{\psi}-1){\overline \psi}i \gamma^{\mu}\partial_{\mu}\psi
-(z_{e}-1)e {\overline \psi}\gamma^{\mu}A_{\mu}\psi \cr
&- (z_{m}-1)m{\overline \psi}\psi
-{1\over 2\alpha}(z_{\alpha}-1)(\partial^{\mu}A_{\mu})^{2}
-{\beta^{2}\over 2 \alpha}(\rho^{2}z_{\alpha}-1)A_{\mu}A^{\mu}.
&(5.10)\cr}
$$
In the equation for ${\cal L}_{\rm ph}$ the parameters $e,m$ and
$$
m_{\gamma}^{2} \equiv {\beta^{2}\over \alpha}
\eqno (5.11)
$$
should be fixed by experiment. This simple equation is one
of the most fundamental in our paper, and it tells us that
the physical mass parameter $m_{\gamma}$ is not $\beta$ but
actually the ratio ${\beta \over \sqrt{\alpha}}$. At
non-perturbative level, $\alpha$ and $\beta$ are independent,
but upon implementing perturbative renormalization we end up
by having a freely specifiable gauge parameter $\alpha$ (as
expected) and a physical parameter $m_{\gamma}$ which is
fixed on observational ground, so that $\beta$ disappears
eventually as an independent parameter,
being equal to $m_{\gamma}\sqrt{\alpha}$.
Note also that, if
$$
\rho={1\over \sqrt{z_{\alpha}}},
\eqno (5.12)
$$
the counterterm Lagrangian reduces to the familiar form in
the Lorenz gauge,$^{(18)}$ and the renormalization of $\beta$
is not independent of the renormalization of $\alpha$, in
agreement with the definition (5.11).
We shall therefore assume that Eq. (5.12) holds from now on,
so that there are no sources of lack of gauge invariance in
${\cal L}_{ct}$ apart from the term arising from the gauge
fixing. Our approach to QED shows that the counterterm
$A_{\mu}A^{\mu}$, which is compatible with Lorentz and charge
conjugation invariance,$^{(18)}$ can indeed be obtained from the
gauge-fixed Lagrangian, while gauge invariance forces us to
weight it with vanishing coefficient (otherwise gauge
invariance would be broken by the quantum dynamics of
QED$^{(18)}$).

We also find that, for arbitrary values of $\alpha_{B}$ and
$\beta_{B}$, the symbol
of the gauge-field operator in QED reads (cf. Sec. 4)
$$
\sigma(P_{\mu \nu})=\left(k^{2}+{m_{\gamma}^{2}\over z_{A}}\right)
g_{\mu \nu}+\left({1\over \alpha_{B}}-1 \right)k_{\mu}k_{\nu}
\equiv \sigma_{\mu \nu}(k),
\eqno (5.13)
$$
since
$$
{\beta_{B}^{2}\over \alpha_{B}}=z_{\alpha}\rho^{2}{1\over z_{A}}
{\beta^{2}\over \alpha}={m_{\gamma}^{2}\over z_{A}},
\eqno (5.14)
$$
by virtue of (5.11) and (5.12).
Its inverse ${\widetilde \sigma}$ is a combination of $g^{\mu \nu}$
and $k^{\mu}k^{\nu}$ with coefficients $\cal A$ and $\cal B$,
respectively, determined from the condition
$$
\sigma_{\mu \nu} {\widetilde \sigma}^{\nu \lambda}
=\delta_{\mu}^{\; \lambda},
\eqno (5.15)
$$
which implies
$$
{\cal A}=\left(k^{2}+{\widetilde m}_{\gamma}^{2}\right)^{-1},
\eqno (5.16)
$$
$$
{\cal B}={(\alpha_{B}-1)\over (k^{2}+\alpha_{B}
{\widetilde m}_{\gamma}^{2})
\left(k^{2}+{\widetilde m}_{\gamma}^{2}\right)},
\eqno (5.17)
$$
upon defining
$$
{\widetilde m}_{\gamma}^{2} \equiv {m_{\gamma}^{2}\over z_{A}}.
\eqno (5.18)
$$
At this stage, the bare photon propagator reads
$$
\bigtriangleup^{\mu \nu}(x,y)=\int {d^{4}k \over (2\pi)^{4}}
\left[{g^{\mu \nu}\over \left(k^{2}+{\widetilde m}_{\gamma}^{2}
\right)}+{(\alpha_{B}-1)k^{\mu}k^{\nu}\over
(k^{2}+\alpha_{B} {\widetilde m}_{\gamma}^{2})
\left(k^{2}+{\widetilde m}_{\gamma}^{2}\right)}\right]
e^{ik \cdot (x-y)}.
\eqno (5.19)
$$
We might have expressed Eq. (5.19) through the bare parameters
$\alpha_{B}$ and $\beta_{B}$ only, but the explicit occurrence
of the physical parameter ${\widetilde m}_{\gamma}^{2}$ will
prove useful in the next section.

Equation (5.19) shows a very important property: the photon
propagator in momentum space falls off like $k^{-2}$ at large $k$
(the same occurs in the Stueckelberg model$^{(19)}$),
which is the behaviour necessary to ensure perturbative
renormalizability.$^{(18)}$ As far as we can see, the deeper roots
of our explicit result (5.19) lie in the use of the BRST method
which is known to lead to perturbative renormalizability
independently of the particular choice of gauge-fixing
condition,$^{(15,16)}$ and also in the preservation of the linear
nature of the gauge-field operator acting on $A_{\mu}$ in
the quantum theory (so that the rescaling (5.1) proves as
useful as in the Lorenz gauge). Any addition by hand of mass
terms to the original gauge-invariant Lagrangian leads instead
to a photon propagator with no regular massless limit,$^{(18)}$
since the purely Maxwell part of the Lagrangian does not lead
to an invertible operator on $A_{\mu}$.

In the massive QED model known so far
in the literature one deals instead with the field equations$^{(19)}$
$$
(i \gamma^{\mu}\partial_{\mu}-M I)\psi=e A^{\mu}\gamma_{\mu}\psi ,
\eqno (5.20)
$$
$$
\partial^{\mu}F_{\mu \nu}+m^{2}A_{\nu}=-e {\overline \psi}
\gamma_{\nu}\psi,
\eqno (5.21)
$$
leading to the photon propagator$^{(19)}$
$$
\bigtriangleup^{\mu \nu}(x,y)=\int {d^{4}k \over (2\pi)^{4}}
\left(g^{\mu \nu}-{k^{\mu}k^{\nu}\over m^{2}}\right)
{-i \over (k^{2}-m^{2}+i \varepsilon)}e^{ik \cdot (x-y)}.
\eqno (5.22)
$$
Here the integrand is constant at large $k$, and this leads to a
non-renormalizable theory, in which the divergence of a Feynman
diagram increases with the number of internal photon lines.
A way out is provided by the introduction of an auxiliary vector
field with free propagator in momentum space given by
$$
-i \left[{g^{\mu \nu}\over (k^{2}-m^{2}+i \varepsilon)}
-{k^{\mu}k^{\nu}\over m^{2}}
\left({1\over (k^{2}-m^{2}+i \varepsilon)}
-{1\over (k^{2}-m_{0}^{2}+i \varepsilon)}\right)\right].
$$
A renormalizable model is now achieved in perturbation theory,
but at a price: the Green functions depend on $m_{0}^{2}$ and
describe an indefinite-metric Hilbert space with ghost particles.$^{(19)}$
Such unpleasant features, however, are not shared by our model,
where mass is not added by hand to the original Lagrangian,
and the photon mass parameter is the physical mass, which is set
to zero {\it at the end of all calculations} on observational ground.
\vskip 2cm
\leftline {\bf 6. RADIATIVE CORRECTIONS IN QED}
\vskip 1cm
A crucial step in quantum electrodynamics is the analysis of
radiative corrections. For our purposes,
we focus on the renormalized
photon propagator with the associated polarization tensor,
since they can be used to evaluate functions of physical interest,
leading in turn to measurable predictions. At a deeper level, we
are aiming to provide direct evidence that physical observables
are independent of the $\beta$-parameter, as will be shown below.

We have just seen that, in the bare theory,
one deals with the gauge-field
operator $P_{\mu \nu}$ and its symbol $\sigma_{\mu \nu}$ with
inverse ${\widetilde \sigma}^{\mu \nu}$. Integration of the
latter in momentum space yields the photon propagator according
to Eq. (5.19). Upon taking into account radiative corrections,
${\widetilde \sigma}^{\mu \nu}(k)$ is replaced by
${\widetilde \Sigma}^{\mu \nu}(k)$ according to the equation$^{(20)}$
$$
{\widetilde \Sigma}^{\mu \nu}(k)={\widetilde \sigma}^{\mu \nu}(k)
+{\widetilde \sigma}^{\mu \lambda}\Pi_{\lambda \rho}
{\widetilde \Sigma}^{\rho \nu}(k),
\eqno (6.1)
$$
where $\Pi_{\lambda \rho}$ is a symmetric rank-two tensor called
the polarization, which is a sum of all diagrams that cannot be
disconnected by cutting only one photon line.
The tensor ${\widetilde \Sigma}^{\mu \nu}$ is
the inverse of the symmetric tensor here denoted by
$\Sigma_{\mu \nu}$. Thus, Eq. (6.1) yields
$$
{\widetilde \Sigma}^{\mu \nu}\Sigma_{\nu \sigma}
=\delta_{\; \sigma}^{\mu}={\widetilde \sigma}^{\mu \nu}
\Sigma_{\nu \sigma}+{\widetilde \sigma}^{\mu \lambda}
\Pi_{\lambda \sigma},
$$
and eventually
$$
\sigma_{\gamma \mu}\delta_{\; \sigma}^{\mu}
=\sigma_{\gamma \mu}{\widetilde \sigma}^{\mu \nu}
\Sigma_{\nu \sigma}+\sigma_{\gamma \mu}
{\widetilde \sigma}^{\mu \lambda}\Pi_{\lambda \sigma},
$$
i.e.
$$
\Pi_{\mu \nu}(k)=\sigma_{\mu \nu}(k)-\Sigma_{\mu \nu}(k).
\eqno (6.2)
$$
The as yet unknown tensors $\Sigma_{\mu \nu}$ and
${\widetilde \Sigma}^{\mu \nu}$ have the general form$^{(20)}$
$$
\Sigma_{\mu \nu}(k)=g_{\mu \nu}u_{1}(k^{2})+k_{\mu}k_{\nu}
u_{2}(k^{2}),
\eqno (6.3)
$$
$$
{\widetilde \Sigma}^{\mu \nu}(k)=g^{\mu \nu}d_{1}(k^{2})
+k^{\mu}k^{\nu}d_{2}(k^{2}),
\eqno (6.4)
$$
and the condition
$$
\Sigma_{\mu \nu}{\widetilde \Sigma}^{\nu \lambda}
=\delta_{\mu}^{\; \lambda}
\eqno (6.5)
$$
yields
$$
d_{1}={1\over u_{1}},
\eqno (6.6)
$$
$$
d_{2}=-{u_{2}d_{1}\over (u_{1}+k^{2}u_{2})}.
\eqno (6.7)
$$
Equation (5.13) for $\sigma_{\mu \nu}(k)$, jointly with (6.2) and
(6.3), yields
$$
\Pi_{\mu \nu}(k)=
g_{\mu \nu}\left(k^{2}+{\widetilde m}_{\gamma}^{2}
-u_{1}\right)+k_{\mu}k_{\nu}\left({1\over \alpha_{B}}-1-u_{2}\right)
= g_{\mu \nu}a_{1}(k^{2})+k_{\mu}k_{\nu}a_{2}(k^{2}).
\eqno (6.8)
$$
Moreover, the condition from current conservation that
$\Pi_{\mu \nu}$ should be transverse, i.e. $k^{\mu}\Pi_{\mu \nu}=0$,
yields $a_{1}=-k^{2}a_{2}$, which leads to
$$
u_{1}+k^{2}u_{2}={1\over \alpha_{B}}
(k^{2}+\alpha_{B} {\widetilde m}_{\gamma}^{2}),
\eqno (6.9)
$$
and hence
$$
\Pi_{\mu \nu}(k)=\left(g_{\mu \nu}-{k_{\mu}k_{\nu}\over k^{2}}
\right)\left(k^{2}+{\widetilde m}_{\gamma}^{2}-u_{1}\right).
\eqno (6.10)
$$
To express that $\Pi_{\mu \nu}$ is gauge-independent we
require that, for some function $f$ of $k^{2}$
which cannot grow faster than $k^{2}$ at large $k$,$^{(20)}$ one has
$$
u_{1}={\widetilde m}_{\gamma}^{2}+f(k^{2}),
\eqno (6.11)
$$
which implies from (6.9) that
$$
u_{2}={1\over \alpha_{B}}-{f(k^{2})\over k^{2}}.
\eqno (6.12)
$$
In the bare theory, $f(k^{2})$ reduces to $k^{2}$, and we find
the bare tensor $\sigma_{\mu \nu}$, here re-written in the
convenient form (cf. Eq. (5.13))
$$
\sigma_{\mu \nu}(k)=\left(g_{\mu \nu}-{k_{\mu}k_{\nu}\over k^{2}}
\right)\left(k^{2}+{\widetilde m}_{\gamma}^{2} \right)
+{k_{\mu}k_{\nu}\over k^{2}}
{1\over \alpha_{B}}(k^{2}+\alpha_{B} {\widetilde m}_{\gamma}^{2}),
\eqno (6.13)
$$
and the radiatively corrected tensor
$$
\Sigma_{\mu \nu}(k)=\left(g_{\mu \nu}-{k_{\mu}k_{\nu}\over k^{2}}
\right)\left(f(k^{2})+{\widetilde m}_{\gamma}^{2} \right)
+{k_{\mu}k_{\nu}\over k^{2}}{1\over \alpha_{B}}
(k^{2}+ \alpha_{B} {\widetilde m}_{\gamma}^{2}).
\eqno (6.14)
$$
Thus, the coefficient of the longitudinal part
${k_{\mu}k_{\nu}\over k^{2}}$ is the same in the bare as well as
in the full theory$^{(20)}$ in agreement with the Ward identity,$^{(18)}$
while the coefficients of the transverse
part $g_{\mu \nu}-{k_{\mu}k_{\nu}\over k^{2}}$ depend on $\alpha_{B}$
and $\beta_{B}$ in such a way that the difference $\sigma_{\mu \nu}(k)
-\Sigma_{\mu \nu}(k)$ is indeed gauge-independent:
$$
\Pi_{\mu \nu}(k)=\sigma_{\mu \nu}(k)-\Sigma_{\mu \nu}(k)
=\left(g_{\mu \nu}-{k_{\mu}k_{\nu}\over k^{2}}\right)
(k^{2}-f(k^{2})).
\eqno (6.15)
$$
Eventually, Eqs. (6.6), (6.7), (6.11) and (6.12) yield the
renormalized photon propagator by integrating in momentum space
the tensor
$$
{\widetilde \Sigma}^{\mu \nu}(k)={g^{\mu \nu}\over
\left(f(k^{2})+{\widetilde m}_{\gamma}^{2}\right)}
+{\left(\alpha_{B} {f(k^{2})\over k^{2}}-1 \right)k^{\mu}k^{\nu}
\over (k^{2}+\alpha_{B} {\widetilde m}_{\gamma}^{2})
\left(f(k^{2})+{\widetilde m}_{\gamma}^{2}\right)}.
\eqno (6.16)
$$

A first application of these formulae is given by radiative
corrections to Coulomb's law, which result from the polarization
of the vacuum around a point charge. Indeed, the ``effective''
potential associated with a charge $q$ takes, in momentum space,
the form$^{(20)}$
$$
{\cal A}^{0}=A^{0}+{\widetilde \Sigma}^{0 \rho}
\Pi_{\rho \lambda}A^{\lambda},
\eqno (6.17)
$$
where $A^{0}$ is proportional to ${q\over k^{2}}$, while the
contraction ${\widetilde \Sigma}^{0 \rho}\Pi_{\rho \lambda}$
reads, from our previous formulae,
$$
{\widetilde \Sigma}^{0 \rho}\Pi_{\rho \lambda}
=\left(\delta_{\; \lambda}^{0}-{k^{0}k_{\lambda}\over k^{2}}
\right){(k^{2}-f(k^{2}))\over \left(f(k^{2})
+{\widetilde m}_{\gamma}^{2}\right)},
\eqno (6.18)
$$
since $k^{0}k^{\rho}d_{2}(k^{2})\Pi_{\rho \lambda}=0$. The
renormalized potential is therefore
$$
{\cal A}^{0}=A^{0}+{(k^{2}-f(k^{2}))\over
\left(f(k^{2})+{\widetilde m}_{\gamma}^{2}\right)}
\left(A^{0}-{k^{0}\over k^{2}}k_{\lambda}A^{\lambda}\right).
\eqno (6.19)
$$
Note that the classical long-range part ${q\over k^{2}}$ resulting from
$A^{0}$ is still present, and eventually
our ${\widetilde m}_{\gamma}$ is set to zero on observational ground.

By virtue of the transverse nature of the polarization tensor
$\Pi_{\rho \lambda}$, the full potential ${\cal A}^{0}$ depends
on gauge parameters $\alpha, \beta$ not
separately, which would have led to unavoidable gauge dependence
(since $\beta =m_{\gamma}\sqrt{\alpha}$), but only through the
combination ${1\over z_{A}}{\beta^{2}\over \alpha}$.
The latter is proportional to
the photon mass parameter $m_{\gamma}^{2}$ in the physical
Lagrangian of perturbative renormalization. {\it Thus, the resulting
short-range potential only depends on a mass parameter in the
physical Lagrangian and is therefore, with the above understanding,
gauge independent} (and so are the coefficients of the transverse
parts in Eqs. (6.13) and (6.14)). An example of gauge
dependence is instead
provided by the coefficient of $k^{\mu}k^{\nu}$ in Eq. (6.16)
for the renormalized photon propagator in momentum space, where the
numerator is equal to $\alpha_{B} {f(k^{2})\over k^{2}}-1$, and the
denominator contains the term
$k^{2}+\beta_{B}^{2}=
k^{2}+\alpha_{B} {\widetilde m}_{\gamma}^{2}$. But such a
coefficient does not affect the renormalized potential, because
Eq. (6.18) holds by virtue of the transverse nature of the
polarization tensor.

Note also that, in the light of previous remarks, the most
general form of Eq. (6.11) is
$$
u_{1}=u_{1} \left(k^{2};{\beta_{B}^{2}\over \alpha_{B}}\right)
=u_{1}(k^{2};{\widetilde m}_{\gamma}^{2}),
\eqno (6.20)
$$
leading to $d_{1}={1\over u_{1}(k^{2};{\widetilde m}_{\gamma}^{2})}$
and
$$
d_{2}={1\over k^{2}}\left[{\alpha_{B} \over
(k^{2}+\alpha_{B} {\widetilde m}_{\gamma}^{2})}
-{1\over u_{1}(k^{2}; {\widetilde m}_{\gamma}^{2})}\right]
\eqno (6.21)
$$
in Eq. (6.4). Once more, only $d_{2}$ is gauge-dependent, since
its first term depends on $\alpha_{B}$.
The renormalized potential (6.19) can be therefore expressed in
the general form
$$
{\cal A}^{0}=A^{0}+{(k^{2}+{\widetilde m}_{\gamma}^{2}
-u_{1}(k^{2};{\widetilde m}_{\gamma}^{2}))\over
u_{1}(k^{2};{\widetilde m}_{\gamma}^{2})}
\left(A^{0}-{k^{0}\over k^{2}}k_{\lambda}A^{\lambda}\right).
\eqno (6.22)
$$
\vskip 2cm
\leftline {\bf 7. MASS TERMS IN NON-ABELIAN THEORY}
\vskip 1cm
The aim of the previous analysis was not to find an alternative
to the Higgs mechanism for Abelian theories, for which no such
mechanism is needed, but rather to prepare the ground for studying
gauge theories relying upon non-Abelian groups.
We can now work out how the above ideas can be applied to a
non-Abelian gauge theory without Higgs field
with group $SU(2) \times U(1)$; an
intriguing theoretical structure will be found to emerge, eventually.

First, the matrix (3.1) is replaced by an
equation representing $4$ of them (here no summation over the index $a$ is
understood)
$$
\Phi^{a} \equiv (I \partial^{\mu}+\beta_{a}\gamma^{\mu})W_{\mu}^{a}.
\eqno (7.1)
$$
Note that, after the experience acquired in
Secs. 3 and 4, we do not write
explicitly matrix indices for $I$ and $\gamma^{\mu}$.
According to the Faddeev-Popov$^{(21)}$
path-integral prescription in the non-Abelian
case, the resulting gauge-averaging term in the Lagrangian is
written with the help of $\Phi^{a}$ and of an invertible
symmetric matrix ${I\over 4} \tau_{ab}$ in the
form (here summation over repeated indices $a$ and $b=0,1,2,3$ is
instead understood)
$$
{1\over 2}\Phi^{a} \; {I\over 4} \; \tau_{ab} \; \Phi^{b}.
$$
>From now on we focus on the resulting mass-like term, which is our main goal.
This reads, summing over all values of $\mu,\nu$ and $a,b$, and taking the
matrix traces as in (3.3),
$$ \eqalignno{
\; &  {1\over 2}g^{\mu \nu} \beta_{a}W_{\mu}^{a}\tau_{ab}
\beta_{b}W_{\nu}^{b} \cr
&= f_{1}(W_{1},W_{2})+f_{2}(Z,Z)+f_{3}(A,A)+f_{4}(Z,A)
+f_{5}(W_{1},W_{2},Z,A).
&(7.2)\cr}
$$
With our notation, where we denote by $k,l$ the indices $a,b$ when taking
the values $1,2$, we find, upon choosing
$$
\tau_{kl}=\tau \; \delta_{kl},
\eqno (7.3)
$$
the formula
$$
f_{1}(W_{1},W_{2}) ={1\over 2}g^{\mu \nu} \beta_{k}W_{\mu}^{k}
\tau_{kl}\beta_{l}W_{\nu}^{l}
={1\over 2}\tau \beta_{k}^{2}W_{\mu}^{k}W_{k}^{\mu}.
\eqno (7.4)
$$
The assumption (7.3) has been made since in the electroweak theory with Higgs
field the mass term associated with $W$ bosons can indeed be cast in
the form (7.4). Moreover, by exploiting the identity
$
g^{\mu \nu} Z_{\mu}Z_{\nu}
=Z_{\mu}Z^{\mu},
$
and choosing
$$
\tau_{03}=\tau_{30},
\eqno (7.5)
$$
we find, by virtue of (1.3),
$$
f_{2}(Z,Z)={1\over 2}\Bigr(\beta_{0}^{2}\sin^{2}\theta \tau_{00}
+\beta_{3}^{2}\cos^{2}\theta \tau_{33}
-2 \beta_{0}\beta_{3}\sin \theta \cos \theta \tau_{03}\Bigr)
Z_{\mu}Z^{\mu},
\eqno (7.6)
$$
where we have set $\theta_{w} \equiv \theta$ for simplicity of notation.
Similarly, we find
$$
f_{3}(A,A)={1\over 2}\Bigr(\beta_{0}^{2}\cos^{2}\theta \tau_{00}
+2\beta_{0}\beta_{3}\sin \theta \cos \theta \tau_{03}
+\beta_{3}^{2}\sin^{2}\theta \tau_{33}\Bigr)A_{\mu}A^{\mu}.
\eqno (7.7)
$$
Furthermore, by virtue of (7.5) and of the identity
$
g^{\mu \nu} (Z_{\mu}A_{\nu}+A_{\mu}Z_{\nu})
=Z_{\mu}A^{\mu}+A_{\mu}Z^{\mu},
$
the mixed term $f_{4}(Z,A)$ reads
$$ \eqalignno{
f_{4}(Z,A)&=
{1\over 2}\Bigr(\beta_{3}^{2}\tau_{33}\sin \theta \cos \theta
-\beta_{0}^{2}\tau_{00}\sin \theta \cos \theta
+\beta_{0} \beta_{3}\tau_{03}
(\cos^{2}\theta -\sin^{2}\theta) \Bigr) \cr
& \times  \Bigr(Z_{\mu}A^{\mu}+A_{\mu}Z^{\mu}\Bigr).
&(7.8)\cr}
$$
Last, the mixed term $f_{5}(W_{1},W_{2},Z,A)$ takes the form
$$ \eqalignno{
\; & f_{5}(W_{1},W_{2},Z,A)={1\over 2}g^{\mu \nu} \Bigr[\beta_{0}
(-\sin \theta Z_{\mu}+\cos \theta A_{\mu})
\tau_{0k}\beta_{k}W_{\nu}^{k} \cr
&+ \beta_{k}W_{\mu}^{k}\tau_{k0}\beta_{0}
(-\sin \theta Z_{\nu}+\cos \theta A_{\nu})
+ \beta_{k}W_{\mu}^{k}\tau_{k3}
\beta_{3}(\cos \theta Z_{\nu}+\sin \theta A_{\nu}) \cr
&+ \beta_{3}(\cos \theta Z_{\mu}
+\sin \theta A_{\mu})\tau_{3k}\beta_{k}W_{\nu}^{k}\Bigr].
&(7.9)\cr}
$$

Now we point out that, since the mixed term $f_{5}$ is not observed in
nature, we have to set
$$
\tau_{0k}=\tau_{k0}=0, \;
\tau_{3k}=\tau_{k3}=0, \forall k=1,2.
\eqno (7.10)
$$
As far as the photon mass $m_{\gamma}$ is concerned, we keep it alive for
the time being because the following calculations will show that it plays
a key role in ensuring internal consistency of our model.
Moreover, we bear in mind that no mixed term $f_{4}(Z,A)$ has ever been
found in experiments. By
virtue of (7.7) and (7.8), these two requirements
lead to the equations
$$
m_{\gamma}^{2}= \Bigr(\beta_{0}^{2}\cos^{2}\theta \tau_{00}
+2\beta_{0}\beta_{3}\sin \theta \cos \theta \tau_{03}
+\beta_{3}^{2}\sin^{2}\theta \tau_{33}\Bigr),
\eqno (7.11)
$$
$$
-\beta_{0}^{2}\sin \theta \cos \theta \tau_{00}
+\beta_{0}\beta_{3}(\cos^{2}\theta-\sin^{2}\theta)\tau_{03}
+\beta_{3}^{2}\sin \theta \cos \theta \tau_{33}=0.
\eqno (7.12)
$$
Such formulae imply that
$$
\tau_{03}=\left(-{\beta_{3}\over \beta_{0}}\tau_{33}
+{m_{\gamma}^{2}\over \beta_{0}\beta_{3}}\right)\tan \theta,
\eqno (7.13)
$$
$$
\tau_{00}={\beta_{3}^{2}\over \beta_{0}^{2}}\tau_{33}\tan^{2}\theta
+{m_{\gamma}^{2}\over \beta_{0}^{2}}(1-\tan^{2}\theta).
\eqno (7.14)
$$
So far, the symmetric matrix $\tau_{ab}$ has been cast in the form
$$
\tau_{ab}=\pmatrix{\tau_{00}&0&0& \tau_{03}\cr
0&\tau &0&0 \cr
0&0& \tau & 0 \cr
\tau_{03}&0&0& \tau_{33} \cr}.
\eqno (7.15)
$$
For consistency, we have now to require that $\tau_{ab}$
should be non-singular, i.e.
$$
{\rm det} \; \tau_{ab}=\tau^{2}(\tau_{00}\tau_{33}
-\tau_{03}^{2}) \not = 0.
\eqno (7.16)
$$
Remarkably, the contributions not involving $m_{\gamma}$ cancel each
other exactly in the determinant (7.16), and one finds
$$
\tau_{00}\tau_{33}-\tau_{03}^{2}
={m_{\gamma}^{2}\over \beta_{0}^{2}}
\left({\tau_{33}\over \cos^{2}\theta}
-{m_{\gamma}^{2}\tan^{2}\theta \over \beta_{3}^{2}}\right).
\eqno (7.17)
$$
At this stage, the photon mass
can be therefore very small for all practical purposes
but nevertheless non-vanishing, for our model to be concretely applicable.
The value of $\tau \beta_{k}^{2}$ in Eq. (7.4) is then fixed by requiring
agreement with the experimental value of $m_{W}^{2}$, i.e. (cf. comments
after (5.10))
$$
m_{W}^{2}=\tau \beta_{1}^{2}=\tau \beta_{2}^{2},
\eqno (7.18)
$$
and the insertion of (7.13) and (7.14) into Eq. (7.6) makes it possible
to fix the value of $\tau_{33}\beta_{3}^{2}$ by requiring agreement
with the observed value of $m_{Z}^{2}$, i.e.
$$
m_{Z}^{2} ={1\over \cos^{2} \theta}\Bigr(\tau_{33}\beta_{3}^{2}
-m_{\gamma}^{2}\sin^{2}\theta \Bigr).
\eqno (7.19)
$$
Interestingly, the first term on the right-hand side of Eq. (7.19) depends
on the Weinberg angle exactly as in Eq. (1.4), which relies instead on
the Higgs boson, but we now have a correction resulting from the photon
mass, which should be non-vanishing to ensure invertibility of the matrix
(7.15) as we have seen. The issue of the photon mass is indeed not entirely
settled. After the investigations in the
seventies,$^{(22-24)}$ more recently
high mass photon pairs have been considered at LEP,$^{(25)}$ while in other
branches of modern physics the concept of photon (effective) mass is
intimately related to the possibility of accelerating photons by moving
plasma perturbations.$^{(26)}$ We will see in the following section how
the $m_{\gamma} \rightarrow 0$ limit can be taken.

Note also that, since we only fix by experiment the products
$\tau \beta_{k}^{2}$ and $\tau_{33}\beta_{3}^{2}$, there is a
residual gauge freedom in chosing, for example, non-vanishing values
of $\tau$ and $\tau_{33}$ (see (7.16)), which then determine the
$\beta$-parameters as functions of $m_{W},m_{Z},m_{\gamma},\theta,\tau$
and $\tau_{33}$. This finding is in agreement with the general
path-integral prescription for quantized gauge theories, according to
which the masses of vector mesons should be independent of the particular
invertible matrix $\tau_{ab}$.$^{(7,17)}$
\vskip 2cm
\leftline {\bf 8. THE $m_{\gamma} \rightarrow 0$ LIMIT AND
ITS IMPLICATIONS}
\vskip 1cm
The $m_{\gamma} \rightarrow 0$ limit deserves now a careful
analysis to make sure that our model is viable. For this purpose,
we first revert to quantum electrodynamics, since our
$4 \times 4$ matrix $\tau_{ab}$ of gauge parameters
corresponds to the $1 \times 1$ matrix ${1\over \alpha}$ in
QED and hence the ${\rm det} \; \tau \rightarrow 0$ limit
corresponds to the $\alpha \rightarrow \infty$ singular limit
in QED. If the latter is understood, we understand the former as well.
Indeed, if we first impose that the four-momentum should have
vanishing contraction with the four-current:
$$
k^{\mu}j_{\mu}=0
\eqno (8.1)
$$
by virtue of current conservation, and then take the limit
$$
\alpha \rightarrow \infty, \; {\rm with} \;
m_{\gamma}^{2} \rightarrow 0
\eqno (8.2)
$$
according to experiment, we recover the massless photon
propagator ${g^{\mu \nu}\over f(k^{2})}$ in momentum space
(cf. (6.16)).

The above order in which the operations are performed is
crucial: first impose Eq. (8.1),
which shows that $k^{\mu}k^{\nu}$ terms do not affect physics
and can be eventually omitted from the integrand defining
the photon propagator. Then take the $\alpha \rightarrow
\infty$ limit while making sure that $m_{\gamma}^{2}$
approaches zero to agree with experiment. {\it With this
understanding}, we can {\it eventually} set to zero
$m_{\gamma}$ in the mass formulae for $SU(2) \times U(1)$ gauge
theory, hence finding
$$
\lim_{m_{\gamma} \to 0}
{m_{Z}^{2}\over m_{W}^{2}}={1\over \cos^{2}\theta}
{\tau_{33}\over \tau}\left({\beta_{3}\over \beta_{1}}
\right)^{2},
\eqno (8.3)
$$
where there exist infinitely many ways of making sure that
$$
{\tau_{33}\over \tau}\left({\beta_{3}\over \beta_{1}}
\right)^{2}=1,
\eqno (8.4)
$$
so that full agreement with the standard formula for
${m_{Z}^{2}\over m_{W}^{2}}$ is eventually recovered.

On setting $m_{\gamma}=0$, the gauge-field operator acting
on $A_{\mu}$ in our version of non-Abelian gauge theory receives
contributions from
$$
{1\over 4}\Bigr(\partial^{\mu}W^{3\nu}
-\partial^{\nu}W^{3\mu}\Bigr)
\Bigr(\partial_{\mu}W_{\nu}^{3}-\partial_{\nu}
W_{\mu}^{3}\Bigr),
$$
$$
{1\over 4}\Bigr(\partial^{\mu}W^{0 \nu}
-\partial^{\nu}W^{0 \mu}\Bigr)
\Bigr(\partial_{\mu}W_{\nu}^{0}-\partial_{\nu}
W_{\mu}^{0}\Bigr),
$$
and from the gauge-averaging term
${1\over 2}\Phi^{a}{I\over 4}\tau_{ab}\Phi^{b}$. Adding
together the three resulting contributions one gets, by
virtue of (1.3) and (7.1), the following photon contribution
to the Lagrangian density: ${\varepsilon \over 2}A^{\nu}
{\cal P}_{\nu}^{\mu}A_{\mu}$, where
$$
{\cal P}_{\nu}^{\; \mu}=-\delta_{\nu}^{\; \mu}
\partial^{\rho}\partial_{\rho}+\left(1- \tau_{33}
\sin^{2}\theta \left({\beta_{3}\over \beta_{0}}-1 \right)^{2}
\right)\partial_{\nu}\partial^{\mu},
\eqno (8.5)
$$
and $\varepsilon=1$ for Euclidean theory, while $\varepsilon=-1$
in Minkowski space-time. At this stage, the gauge-field operator
for photons can be reduced to the minimal form
$-\delta_{\nu}^{\; \mu}\partial^{\rho}\partial_{\rho}$ by
choosing the ratio of gauge parameters in such a way that
$$
{\beta_{3}\over \beta_{0}}=1+{1\over \sqrt{\tau_{33}}}
{1\over \sin \theta}.
\eqno (8.6)
$$

Similarly to Eqs. (5.5) and (5.12), gauge parameters can be
renormalized by setting
$$
(\tau_{ab})_{B}=f_{ab} \; \tau_{ab},
\eqno (8.7)
$$
$$
(\beta_{a})_{B}=\rho_{a} \; \beta_{a},
\eqno (8.8)
$$
and requiring that the counterterm Lagrangian should contain
no massive term for photons. This leads to the equations
$$
\rho_{0}^{2}f_{00}=1,
\eqno (8.9)
$$
$$
\rho_{0}\rho_{3}f_{03}=1,
\eqno (8.10)
$$
$$
\rho_{3}^{2}f_{33}=1,
\eqno (8.11)
$$
which are solved by
$$
\rho_{0}={1\over \sqrt{f_{00}}},
\eqno (8.12)
$$
$$
\rho_{3}={\sqrt{f_{00}}\over f_{03}}
={1\over \sqrt{f_{33}}}.
\eqno (8.13)
$$

It should however be stressed that,
since there is no need to perform a gauge averaging in the path
integral for spin-${1\over 2}$ fields,$^{(7,17)}$ we do not succeed
in finding alternatives to the Higgs mechanism for the generation
of fermionic masses.
\vskip 2cm
\leftline {\bf 9. CONCLUDING REMARKS}
\vskip 1cm
The current models of mass generation in field theory rely on the
assumption that Higgs bosons exist,
with the associated Higgs mechanism.$^{(1)}$
However, if Higgs bosons were to remain elusive,
the problem remains to understand to which extent the general principles
of quantum field theory make it possible to account for the existence of
massive vector bosons.

Indeed, when Higgs elaborated his model$^{(1)}$ the emphasis was very
much on gauge-invariant Lagrangians, whereas it is by now clear
and well accepted that the starting point for quantization of gauge
theories is a Lagrangian (still called classical in Ref. 16)
consisting of three ingredients: a gauge-invariant part, a
gauge-breaking term and contribution of ghost fields.$^{(7)}$ If, for
a moment, we no longer assume that Higgs fields exist, we still
have to work within this broader framework, although it is by no
means obvious that we are going to find a viable scheme
even just for describing the occurrence of mass.
As a matter of fact, we have not succeeded in this
respect, but our stronger findings, of field-theoretical interest, are
as follows.
\vskip 0.3cm
\noindent
(i) Supplementary condition in QED chosen in the non-linear form
(3.4). This suggests that $\gamma$-matrices generate the
matrix $\Phi_{i}^{\; j}$ in (3.1) which, in turn, acts
as a `potential' for the gauge-fixing functional through Eq. (3.4).
\vskip 0.3cm
\noindent
(ii) New photon propagators in quantum electrodynamics, with a
possibly deeper perspective on the massless nature of
photons in vacuum QED (Secs. 4 and 5).
\vskip 0.3cm
\noindent
(iii) Renormalization of the gauge parameter $\beta$ in such a way that
the counterterm Lagrangian has vanishing coefficient of
$A_{\mu}A^{\mu}$ (Sec. 5), as in the ordinary formulation of
QED in linear covariant gauges.$^{(18)}$
\vskip 0.3cm
\noindent
(iv) Evaluation of the renormalized photon propagator in our gauges,
and proof of gauge independence of the associated short-range potential,
adding evidence in favour of our model being physically relevant.

We have also considered mass terms in $SU(2) \times U(1)$ gauge theory,
generalizing the Maxwell construction with the help of the invertible matrix
(7.15), with gauge-averaging term reading
$$
{1\over 2}(\Phi^{a})_{i}^{\; j}(\Omega_{ab})_{j}^{\; k}
(\Phi^{b})_{k}^{\; i},
$$
where $(\Omega_{ab})_{j}^{\; k}$ is the $16 \times 16$
matrix having diagonal form
$$
(\Omega_{ab})_{j}^{\; k} \equiv {1\over 4}
\delta_{j}^{\; k}\tau_{ab}
={1\over 4}\pmatrix{\tau_{ab} & 0 & 0 & 0 \cr
0 & \tau_{ab} & 0 & 0 \cr
0 & 0 & \tau_{ab} & 0 \cr
0 & 0 & 0 & \tau_{ab} \cr}.
\eqno (9.1)
$$
Note that both $\tau_{00}$ and $\tau_{03}$ depend linearly
on $\tau_{33}$, and hence {\it there exist infinitely many choices of}
$\tau_{33}$ {\it leading to the same values of} $m_{Z}^{2}$ (see (7.19)),
and similarly for $\tau$ and $m_{W}^{2}$ (see (7.18)). Moreover, Eqs.
(7.18) and (7.19) make it possible to express $\beta_{1},\beta_{2}$ and
$\beta_{3}$ in terms of $m_{w},\tau,m_{\gamma},
m_{Z},\tau_{33}$ and the Weinberg angle
$\theta$. Thus, eventually, the gauge-averaging functional is only found
up to infinitely many possible choices of $\tau$ and $\tau_{33}$, in
agreement with the basic requirement
that {\it masses of vector mesons should
be independent of both} $\Phi^{a}$ {\it and} $\tau_{ab}$.$^{(7,17)}$
The associated massless limit for photons has been studied in
detail in Sec. 8; this is a singular limit which can only be taken
after the general path-integral formulae for the photon propagator
have been worked out, as we have shown therein. If our prescription
for taking the $m_{\gamma} \rightarrow 0$ limit in Sec. 8 is
rejected, the content of our paper remains purely Abelian but still
quite interesting: the photon mass parameter is independent of any
particular choice of the gauge parameter $\alpha$ and is set to
zero on observational ground. It remains true that the
gauge-averaging procedure has removed redundant degrees of
freedom while not affecting the physics.

A legitimate objection against our scheme might be that, in the
Abelian case, the mass has physical relevance since it has a
non-trivial cohomological content,$^{(15,16)}$ whereas we might be
hiding this property by appealing to gauge averaging
in the path integral. However, non-trivial cohomology is relevant
for the Abelian Higgs--Kibble model which assumes the existence
of fundamental scalar fields,
whereas we have assumed neither fundamental scalar fields nor
massive photons from the outset.

While our paper
was in preparation, the LEP collaboration has announced data which
can be accounted for by assuming a Higgs boson with mass of about
115 GeV.$^{(27-31)}$ New theoretical investigations have been therefore
performed, including a probability density calculation of the Higgs boson
mass.$^{(32)}$ However, there is not yet conclusive evidence in favour of
the existence of Higgs bosons, and only the
Large Hadron Collider (LHC) can rule out some of the existing models.
In particular, our model reflects the
desire to develop theoretical physics with the minimal amount of structures
and making use of known fields only.

As we acknowledge in Sec. 8, no path-integral approach can
succeed in generating mass terms for spin-${1\over 2}$ fields.
Nevertheless, it appears relevant to have
found a mechanism for dealing with ({\it but not truly generating !})
massive terms in the quantization
of gauge theories while preserving perturbative renormalizability
and independence of particular values of gauge parameters, as we
have done explicitly in the Abelian case and advocated in
non-Abelian gauge theories.

At field-theoretical level, it now appears important
to prove perturbative renormalizability and
investigate the possible occurrence of Gribov ambiguities$^{(33)}$
in our gauges for non-Abelian theories (although our
analysis is inspired by perturbation theory, a framework
where Gribov copies are expected to decouple$^{(18)}$),
as well as mass terms for ghost fields.$^{(16)}$ Moreover, since we
end up by putting the emphasis on the space of four-vectors with
components given by $4 \times 4$ matrices, which is a natural
structure for theories incorporating fermions, a last effort is
in order before ruling out that our scheme might be relevant for
a deeper understanding of the standard model in particle physics.
If this effort fails, one might have to resort to the mechanism
suggested by Gribov in Ref. 34, according to which, in the
standard model without elementary Higgs the fact that the
$U(1)$ coupling becomes of order $1$ at the Landau scale
leads to spontaneous symmetry breaking and generation of masses.
\vskip 1cm
\leftline {\bf ACKNOWLEDGMENTS}
\vskip 1cm
The work of G. Esposito has been partially supported by the Progetto
di Ricerca di Interesse Nazionale {\sl SINTESI 2000}.
He is indebted to Francesco Nicodemi for patiently listening
to his ideas and enlightening comments, to
Giorgio Immirzi and Pietro Santorelli for
conversations and a careful reading of the manuscript,
to Ivan Avramidi and Giuseppe Marmo for introducing him
to the use of symbols, and to Stanislav
Alexeyev for help in checking the mass formulae.
\vskip 10cm
\leftline {\bf REFERENCES}
\vskip 1cm
\noindent
\item {1.}
P. W. Higgs, ``Broken symmetries and the masses of gauge bosons,''
{\it Phys. Rev. Lett.} {\bf 13}, 508 (1964); ``Spontaneous
symmetry breakdown without massless bosons,''
{\it Phys. Rev.} {\bf 145}, 1156 (1966).
\item {2.}
S. Heinemeyer and G. Weiglein, ``Higgs-mass predictions and
electroweak precision observables in the standard model and
the MSSM,'' {\it Nucl. Phys. Proc. Suppl.} {\bf 89}, 216 (2000).
\item {3.}
S. L. Glashow, ``Towards a unified theory: threads in a tapestry,''
{\it Rev. Mod. Phys.} {\bf 52}, 539 (1980).
\item {4.}
A. Salam, ``Grand unification of fundamental forces,''
{\it Rev. Mod. Phys.} {\bf 52}, 525 (1980).
\item {5.}
S. Weinberg, ``Conceptual foundations of the unified theory of
weak and electromagnetic interactions,''
{\it Rev. Mod. Phys.} {\bf 52}, 515 (1980).
\item {6.}
B. S. DeWitt, {\it Dynamical Theory of Groups and Fields}
(Gordon and Breach, New York, 1965).
\item {7.}
B. S. DeWitt, ``The space-time approach to quantum field theory,''
in {\it Relativity, Groups and Topology II}, edited
by B. S. DeWitt and R. L. Stora (North Holland, Amsterdam, 1984).
\item {8.}
I. G. Avramidi and G. Esposito, ``Gauge theories on manifolds with
boundary,'' {\it Commun. Math. Phys.} {\bf 200}, 495 (1999).
\item {9.}
G. Esposito, ``Non-local boundary conditions in Euclidean quantum
gravity,'' {\it Class. Quantum Grav.} {\bf 16}, 1113 (1999).
\item {10.}
G. Esposito, ``New kernels in quantum gravity,''
{\it Class. Quantum Grav.} {\bf 16}, 3999 (1999).
\item {11.}
G. Esposito and C. Stornaiolo, ``Non-locality and ellipticity in a
gauge-invariant quantization,''
{\it Int. J. Mod. Phys. A} {\bf 15}, 449 (2000).
\item {12.}
G. Esposito, ``Boundary operators in quantum field theory,''
{\it Int. J. Mod. Phys. A} {\bf 15}, 4539 (2000).
\item {13.}
L. Lorenz, ``On the identity of the vibrations of light
with electrical currents,''
{\it Phil. Mag.} {\bf 34}, 287 (1867); E. T. Whittaker,
{\it The History of the Theories of Aether and Electricity}
(Longman, London, 1910).
\item {14.}
P. K. Townsend, ``Cosmological constant in supergravity,''
{\it Phys. Rev. D} {\bf 15}, 2802 (1977).
\item {15.}
C. Becchi, A. Rouet, and R. Stora, ``Renormalization of the Abelian
Higgs--Kibble model,''
{\it Commun. Math. Phys.} {\bf 42}, 127 (1975).
\item {16.}
C. Becchi, A. Rouet, and R. Stora, ``Renormalization of gauge
theories,'' {\it Ann. Phys. (N.Y.)} {\bf 98}, 287 (1976).
\item {17.}
B. S. DeWitt, ``Quantum gravity: the new synthesis,''
in {\it General Relativity, an Einstein Centenary Survey},
edited by S. W. Hawking and W. Israel (Cambridge University Press,
Cambridge, 1979).
\item {18.}
J. Zinn--Justin, {\it Quantum Field Theory and Critical
Phenomena} (Clarendon Press, Oxford, 1989);
S. Weinberg, {\it The Quantum Theory of Fields. Vol. I}
(Cambridge University Press, Cambridge, 1996); R. Ticciati,
{\it Quantum Field Theory for Mathematicians} (Cambridge
University Press, Cambridge, 1999).
\item {19.}
J. H. Lowenstein and B. Schroer, ``Gauge invariance and Ward
identities in a massive-vector-meson model,''
{\it Phys. Rev. D} {\bf 6}, 1553 (1972);
C. Itzykson and J. B. Zuber, {\it Quantum Field Theory}
(McGraw--Hill, New York, 1985).
\item {20.}
L. D. Landau, E. M. Lifshitz, and L. P. Pitaevskii,
{\it Relativistic Quantum Theory, Vol. 4 of Course
of Theoretical Physics} (Pergamon Press, Oxford, 1971);
V. N. Gribov and J. Nyiri, {\it Quantum Electrodynamics}
(Cambridge University Press, Cambridge, 2001).
\item {21.}
L. D. Faddeev and V. Popov, ``Feynman diagrams for the Yang--Mills
field,'' {\it Phys. Lett. B} {\bf 25}, 29 (1967).
\item {22.}
A. S. Goldhaber and M. M. Nieto, ``Terrestrial and extraterrestrial
limits on the photon mass,''
{\it Rev. Mod. Phys.} {\bf 43}, 277 (1971).
\item {23.}
J. V. Hollweg, ``Improved limit on photon rest mass,''
{\it Phys. Rev. Lett.} {\bf 32}, 961 (1974).
\item {24.}
L. Davis Jr., A. S. Goldhaber, and M. M. Nieto, ``Limit on the
photon mass deduced from Pioneer-10 observation's of Jupiter's
magnetic field,'' {\it Phys. Rev. Lett.} {\bf 35}, 1402 (1975).
\item {25.}
G. W. Wilson, ``High Mass Photon Pairs at LEP'', in Marseille
1993, High Energy Physics (1993) 285.
\item {26.}
J. T. Mendonca, A. M. Martins, and A. Guerreiro, ``Field
quantization in a plasma: photon mass and charge,''
{\it Phys. Rev. E} {\bf 62}, 2989 (2000).
\item {27.}
P. Abreu et al., ``Search for the standard model Higgs boson at
LEP in the year 2001,''
{\it Phys. Lett. B} {\bf 499}, 23 (2001).
\item {28.}
J. Ellis, G. Ganis, D. V. Nanopoulos and K. A. Olive,
``What if the Higgs boson weighs 115 GeV?,''
{\it Phys. Lett. B} {\bf 502}, 171 (2001).
\item {29.}
M. Acciarri et al., ``Search for neutral Higgs bosons of the
minimal supersymmetric standard model in $e^{+}e^{-}$ interactions
at $\sqrt{s}$=192 GeV-202 GeV,''
{\it Phys. Lett. B} {\bf 503}, 21 (2001).
\item {30.}
M. Acciarri et al., ``Search for the standard model Higgs boson
in $e^{+}e^{-}$ collisions at $\sqrt{s}$ up to 202 GeV,''
{\it Phys. Lett. B} {\bf 508}, 225 (2001).
\item {31.}
P. Achard et al., ``Standard model Higgs boson with the L3
experiment at LEP,'' {\it Phys. Lett. B} {\bf 517}, 319 (2001).
\item {32.}
J. Erler, ``The probability density of the Higgs boson mass,''
{\it Phys. Rev. D} {\bf 63}, 071301 (2001).
\item {33}
V. N. Gribov, ``Quantization of non-Abelian gauge theories,''
{\it Nucl. Phys. B} {\bf 139}, 1 (1978).
\item {34}
V. N. Gribov, ``Higgs and top quark masses in the standard model
without elementary Higgs boson,''
{\it Phys. Lett. B} {\bf 336}, 243 (1994).

\bye